\documentclass[conference]{IEEEtran}
\IEEEoverridecommandlockouts
\usepackage{cite}
\usepackage{amsmath,amssymb,amsfonts,amsthm}
\usepackage{mathtools}
\usepackage{algorithmic}
\usepackage{graphicx}
\usepackage{textcomp}
\usepackage{xcolor}
\usepackage{gensymb}
\def\BibTeX{{\rm B\kern-.05em{\sc i\kern-.025em b}\kern-.08em T\kern-.1667em\lower.7ex\hbox{E}\kern-.125emX}}

\DeclarePairedDelimiter{\round}\lfloor\rceil
\DeclarePairedDelimiter{\lrp}{(}{)}

\newtheorem{proposition}{Proposition}
\newtheorem{lemma}{Lemma}

\newcommand{\rmd}{{\,\mathrm{d}}}
\renewcommand{\IEEEQED}{\IEEEQEDopen}

\begin{document}
\bstctlcite{IEEEexample:BSTcontrol}

\title{Study of Intelligent Reflective Surface Assisted Communications with One-bit Phase Adjustments}

\author{
    \IEEEauthorblockN{Tianxiong Wang\IEEEauthorrefmark{1}, Gaojie~Chen\IEEEauthorrefmark{2}, Justin P.~Coon\IEEEauthorrefmark{1}, and Mihai-Alin~Badiu\IEEEauthorrefmark{1}}
    \IEEEauthorblockA{
        \IEEEauthorrefmark{1}Department of Engineering Science, University of Oxford, Oxford OX1 3PJ, United Kingdom\\
    }
    \IEEEauthorblockA{
        \IEEEauthorrefmark{2}Department of Engineering, University of Leicester, Leicester LE1 7RH, United Kingdom\\
        e-mail: \{tianxiong.wang, justin.coon, mihai.badiu\}@eng.ox.ac.uk, gaojie.chen@leicester.ac.uk
    }
}

\maketitle

\begin{abstract}
We analyse the performance of a communication link assisted by an intelligent reflective surface (IRS) positioned in the far field of both the source and the destination. A direct link between the transmitting and receiving devices is assumed to exist.  Perfect and imperfect phase adjustments at the IRS are considered. For the perfect phase configuration, we derive an approximate expression for the outage probability in closed form. For the imperfect phase configuration, we assume that each element of the IRS has a one-bit phase shifter ($0\degree,\,180\degree$) and an expression for the outage probability is obtained in the form of an integral.  Our formulation admits an exact asymptotic (high SNR) analysis, from which we obtain the diversity orders for systems with and without phase errors. We show these are $N+1$ and $\frac1 2 (N+3)$, respectively.  Numerical results confirm the theoretical analysis and verify that the reported results are more accurate than methods based on the central limit theorem (CLT). 
\end{abstract}


\section{Introduction}

Recently, the idea of using an \textit{intelligent reflective surface} (IRS) located between transmitting and receiving devices to create a ``tunable'' propagation environment has been proposed as a future enhancement to the physical layer of wireless systems~\cite{b4,b1,b3}. An IRS consists of many reflective elements made of electromagnetic (EM) material, which can reconfigure the incident waves for different purposes, such as focusing (reflecting) the waves towards a prescribed target~\cite{zr0} or enhancing security~\cite{RS19}. Compared with other related technologies, such as relaying, IRS-assisted communication offers energy-efficiency, convenient deployment, and a full-band response \cite{b4}. 

Different implementations of IRSs -- including smart reflect-arrays, software-defined hypersurfaces, and frequency-selective surfaces -- have been investigated~\cite{b1,b2,b3}. It has been shown that an IRS can improve  communication performance between the source and the destination if the phase shifts are configured properly~\cite{bm}. In~\cite{zr0}, a beamforming technique was proposed for a multiple-input, single-output (MISO) system in the presence an IRS to maximize the power of the received signals. The authors in~\cite{zr} discussed  beamforming optimization for an IRS-assisted system when the elements of the IRS have discrete phase shifts. In~\cite{b5} and~\cite{b6}, the performance of an IRS-aided communication link was compared with amplify-and-forward (AF) relaying and decode-and-forward (DF) relaying, respectively. In~\cite{b7}, the authors derived the performance bounds of an IRS system without the presence of a direct link by using a central limit theorem (CLT) approximation. As a further step, phase errors were taken into consideration for an IRS system without the direct link in~\cite{b8}. By using the CLT, the authors of that work showed that the channel distribution is Nakagami. It is known that the CLT is inaccurate when the number of elements in the IRS is small; however, it also turns out that the approximation error attributed to the CLT can be significant in the high SNR regime. The source of this inaccuracy results from the fact that the CLT is used to approximate a positive real random variable as a Gaussian, which has infinite support. To circumvent the CLT issue, a gamma distribution was used to model the channel fading of each reflecting path in~\cite{b9}.  This approach leads to more accurate results; however, most recent work that uses the gamma model ignores the direct link and the possibility of phase adjustment errors at the IRS.

In this paper, we shed further light on the performance of IRS-aided systems by analyzing the outage probability and calculating the diversity order of such systems when (1) a direct link is present and (2) the IRS performs a binary phase adjustment at each element.  We also treat the benchmark case where perfect phase adjustments are made. Further, we do not employ a CLT approach. 


\textit{Notation:} The probability density function (PDF) and the cumulative distribution function (CDF) of a random variable \(X\) are denoted by \(f_X(\cdot)\) and \(F_X(\cdot)\). \(\Gamma(a) = \int_0^\infty t^{a-1}e^{-t}\rmd t\) and \(\Gamma(a \, ,x) = \int_x^\infty t^{a-1}e^{-t}\rmd t\) represent the gamma function and the upper incomplete Gamma function. \({K_n}( \cdot )\) denotes the modified Bessel function of the second kind with order \(n\), and $\round{x}$ is the integer closest to $x$. ${\mathbb{P}}(\cdot)$ is the probability operator. $D(\cdot||\cdot)$ and $I(\cdot ; \cdot)$ denote the relative entropy and mutual information functions.

\section{System Model}
The system model considered in this paper includes a source (S) node, a destination (D) node and an IRS with \(N\) reflecting elements operating in the far field of both S and D. Single, half-duplex antennas operate at the source and destination nodes. The reflecting elements, labelled \({\rm R}_n, n\in\{1, ..., N\}\), can scatter the incident waves independently. 

We consider the case where a direct link exists from S to D. It is assumed that the path losses of the S-to-\({\rm R}_n\) channels are the same for all $n$ and the path losses of the \({\rm R}_n\)-to-D channels are also equal for all $n$. Denote the complex attenuation coefficients corresponding the S-to-\({\rm R}_n\), \({\rm R}_n\)-to-D, and S-to-D channels by \(h_{1n}'\), \({{h'_{2n}}}\) and \({{h'_{sd}}}\), respectively. Under this model, the received signal can be written as 
\begin{equation}\small
y' = \left( {\sum\limits_{n = 1}^N {\sqrt {{\xi _1}} {h'_{1n}}{e^{j{\theta _n}}}\sqrt {{\xi _2}} {h'_{2n}} + \sqrt {{\xi _d}} {h'_{sd}}} } \right)\sqrt P x + {w'}
\end{equation}
where \({h'_{1n}}\), \({h'_{2n}}\) and \({h'_{sd}}\) are circularly symmetric, complex Gaussian random variables, each with zero mean and unit variance; \({{\xi _1}}\), \({{\xi _2}}\), and \({{\xi _d}}\) denote the path losses of the S-to-\({\rm R}_n\), \({\rm R}_n\)-to-D, and S-to-D channels, respectively; \({\theta _n}\) is the phase shift induced by the \(n\)th reflecting element at the IRS; \(P\) is the transmit power at S; \(x\) is the information symbol with unit power, i.e., \({\rm{E}}[{\left| x \right|^2}] = 1\); and \(w' \sim \mathcal{CN}(0,{{\sigma_w'}^2})\) represents additive white Gaussian noise (AWGN) at the destination.

Dividing the left-hand and right-hand sides of the equation above by \(\frac{{\sqrt {{\xi _1}{\xi _2}} }}{2}\), we have
\begin{equation} \small
y = \left( {\sum\limits_{n = 1}^N {{h_{1n}}{e^{j{\theta _n}}}{h_{2n}} + {h_{sd}}} } \right)\sqrt P x + w
\label{eqsys}
\end{equation}
where \({{{h}_{1n}}} \sim \mathcal{CN}(0,2)\); \({{{h}_{2n}}} \sim \mathcal{CN}(0,2)\); \({h_{sd}} \sim \mathcal{CN}(0,{\frac{{{4\xi _d}}}{{{\xi _1}{\xi _2}}}})\); and \(w \sim \mathcal{CN}(0,{\sigma_w ^2})\). It follows that \({\left| {{h_{1n}}} \right|}\), \({\left| {{h_{2n}}} \right|}\) and \(\left| {{h_{sd}}} \right|\) are independent, Rayleigh distributed random variables. Furthermore, \({\left| {{h_{1n}}} \right|}\) and \({\left| {{h_{2n}}} \right|}\) have unit scale parameters. We write the scale parameter of \(\left| {{h_{sd}}} \right|\) as \(\sigma_d=\sqrt {\frac{{{2\xi _d}}}{{{\xi _1}{\xi _2}}}} \).  It should be clear that it is assumed that all the channels are flat and slow fading and mutually independent. For clarity, in the rest of the paper, we will use \eqref{eqsys} to analyze the performance of the system.

\subsection{Perfect Phase Alignment}
Ideally, the phase shift of each reflecting element should satisfy \({\theta _n} = \arg ({h_{sd}}) - \arg ({h_{1n}}) - \arg ({h_{2n}})\) to maximize the received SNR~\cite{b6}. In this case, the received signal at D is  
\begin{equation} \small
{y} = H{e^{j\arg ({h_{sd}})}}\sqrt P x + w \label{2.17.1}
\end{equation}
where 
\begin{equation} \small
H = \sum\limits_{n = 1}^N {\left| {{h_{1n}}} \right|\left| {{h_{2n}}} \right| + \left| {{h_{sd}}} \right|} = S + R.\label{2.17.2}
\end{equation}
Thus, the received SNR can be written as
\begin{equation} \small
{\gamma _1} = \frac{{P}}{{\sigma _w^2}}{H^2} = {H^2}{\gamma _t}
\end{equation}
where \({\gamma _t = P/\sigma_w^2}\) is the transmit SNR.

\subsection{One-bit Phase Adjustment}
Although aligning the phases of the reflected paths to the phase of the direct link can optimize the system performance, the phases of the reflecting elements cannot be set precisely in reality due to imperfect channel knowledge and the finite precision inherent in the phase alignment operation.  Here, we focus on the latter issue, which manifests in phase quantization at each element in the IRS.  Mathematically, the resulting phase adjustment at the $n$th element can be written as \({\theta _n} = \arg ({h_{sd}}) - \arg ({h_{1n}}) - \arg ({h_{2n}}) + {\phi _n}\), where \({\phi _n}\) denotes the phase error introduced through quantization at the $n$th element. We assume that each element of the IRS is a one-bit phase shifter, either leaving the phase of the incident wave unaltered or shifting it by $180\degree$. It follows that the phase errors \({\phi _n}\), \(n \in \{1,...,N\}\) are mutually independent and uniformly distributed on the interval~\([ - {\pi }/{2},{\pi }/{2}]\)~\cite[Prop.~1]{zr}.

In this case, the composite channel, which we label \(G\) instead of $H$ for notational clarity, can be  written as
\begin{equation} \small
G = \sum\limits_{n = 1}^N {\left| {{h_{1n}}} \right|\left| {{h_{2n}}} \right|{e^{j{\phi _n}}}}  + \left| {{h_{sd}}} \right| \label{2.17.5}
\end{equation}
and the corresponding received SNR for one-bit phase adjustment can be written as 
\begin{equation} \small
{\gamma _2} = \frac{{P}}{{\sigma _w^2}}{{\left| G \right|}^2}={\left| G \right|^2}{\gamma _t}.
\end{equation}

\section{Outage Probability}
The outage probability is equivalent to the received SNR distribution, i.e., the probability that the received SNR falls below a threshold \({\gamma _{\text{th}}}\):
\begin{equation} \small
    {{{P}}^{(i)}_{\text{out}}}(\gamma _{\text{th}}) = {\mathbb{P}}\left( {{\gamma_i} < {\gamma _{\text{th}}}} \right) = {F_{{\gamma_i}}}\left( {{\gamma _{\text{th}}}} \right)
\label{out1}
\end{equation}
where $i\in\{1,2\}$ denotes the perfect and one-bit phase adjustment scenarios, respectively. The calculation of $P^{(i)}_{\text{out}}$ amounts to computing the CDF of the channel gain associated with $\gamma_i$.  For example, for the case of perfect phase alignment, we have
\begin{equation} \small
    {{P}}^{(1)}_{\text{out}}(\gamma_{\text{th}}) = F_{H}\left(\sqrt{\frac{\gamma_{\text{th}}}{\gamma_t}}\right).
\end{equation}
Hence, in what follows, we treat the calculation of the channel random variables for the cases of perfect and imperfect phase adjustment. 

\subsection{Perfect Phase Alignment}
Let \(H_n = \left| {{h_{1n}}} \right|\left| {{h_{2n}}} \right|\). $H_n$ follows a double Rayleigh distribution and the PDF of \({H_n}\) is given by~\cite{b10}
\begin{equation}\small
f_{{H_n}}(x) = x{K_0}(x). 
\label{eq888}
\end{equation}
Unfortunately, continuing with this form of the calculation will prove fruitless; hence, in accordance with~\cite{b9}, we adopt a good approximation for the PDF of \({H_n}\) by using the gamma distribution:
\begin{equation}\small
    {f_{{H_n}}}(x) \approx \frac{{{x^{k - 1}}{e^{ - \frac{x}{\theta }}}}}{{{\theta ^k}\Gamma (k)}}
\label{2.17.9}
\end{equation}
where \(k = \frac{{{\pi ^2}}}{{16 - {\pi ^2}}}\) and \(\theta = \frac{{16 - {\pi ^2}}}{{2\pi }}\).
Since \({H_n}\) for \(n \in \{1,...,N\}\) are independent and identically distributed (i.i.d.), we can write the PDF of \({S} = \sum_{n = 1}^N {\left| {{h_{1n}}} \right|\left| {{h_{2n}}} \right|} \) as
\begin{equation}\small
    {f_{{S}}}(x) \approx \frac{{{x^{Nk - 1}}{e^{ - \frac{x}{\theta }}}}}{{{\theta ^{Nk}}\Gamma (Nk)}}
\label{2.17.1a}.
\end{equation}
Using this gamma-based framework, we can obtain the following approximation for the CDF of the composite  channel~\(H\).

\begin{proposition}[CDF of $H$]
    The CDF of \(H\) can be approximated as
    \begin{multline}\small
    {F_H}(t) \approx 1 - \frac{\Gamma \left( {{\round{Nk}},\frac{t}{\theta }} \right)}{{\Gamma ({\round{Nk}})}} \nonumber \\
    - A(t)\sum\limits_{i = 0}^{\round{Nk} - 1} \binom{\round{Nk} - 1}{i} B_i \,m(t)^{\round{Nk} - 1 - i} \label{217.2}
    \end{multline}
    where
    \begin{align}\small
    A(t) &= \frac{{\exp \left( {\frac{{\sigma _d^2}}{{2{\theta ^2}}} - \frac{t}{\theta }} \right)}}{{\Gamma \left( {{\round{Nk}}} \right){\theta ^{{\round{Nk}}}}}} \nonumber\\
    {B_i} &=\small{ {2^{\frac{{i - 1}}{2}}}\sigma _d^{i + 1}\left( {\Gamma \left( {\frac{{i + 1}}{2},\frac{{{m^2}}}{{2\sigma _d^2}}} \right) - \Gamma \left( {\frac{{i + 1}}{2},\frac{{\sigma _d^2}}{{2{\theta ^2}}}} \right)} \right)\nonumber }\\
    m(t) &= t - \frac{{\sigma _d^2}}{\theta } \quad\text{and}\quad {\sigma _d}=\sqrt {\frac{{2{\xi _d}}}{{{\xi _1}{\xi _2}}}}. \nonumber
    \end{align}
\end{proposition}
\begin{IEEEproof}
    See the appendix.
\end{IEEEproof}

\begin{figure}[t!]
\centerline{\includegraphics[scale=0.45]{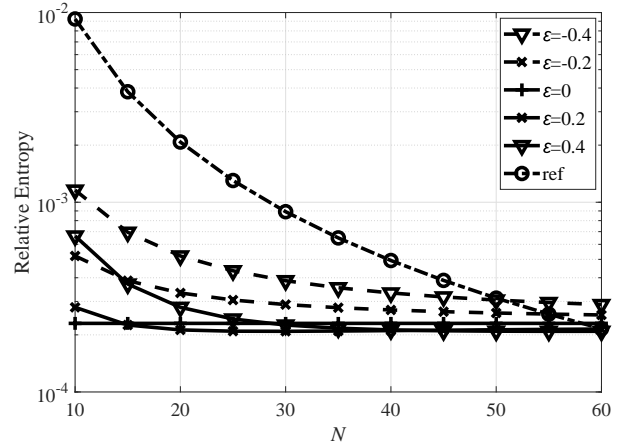}}   
\caption{Relative entropy $D(f_{H_n}||f_{\text{app}})$ versus $N$ for different values of \( \varepsilon\).  The ``ref'' curves corresponds to the relative entropy of Student's $t$-distribution with $N$ degrees of freedom and its normal approximation.}
\label{fig1}
\end{figure}

The approximation in Proposition 1 arises from two aspects of the calculation: (1) the gamma distribution is used to approximate the double Rayleigh distribution, and (2) to obtain a simple, tractable expression, we induce the rounding $\round{Nk}$.  To explore the accuracy of the approximation, we compute the relative entropy of the true distribution and the approximation.  Since \(Nk\) is approximated by \({\round{Nk}}\), the result stated in Proposition 1 is equivalent to that which would be obtained if we used a gamma approximation for $H_n$ with parameters
\begin{equation}\small
k' = \frac{{{\round{Nk}}}}{N} = k + \frac{\varepsilon }{N} \quad {\text{and}} \quad \theta ' = \theta \end{equation}
where \(\varepsilon  \in \left[ {-0.5,0.5} \right]\) is the difference between \(Nk\) and \({\round{Nk}}\). The relative entropy of the exact distribution and the approximate (gamma) distribution in this case is defined as
\begin{equation}\small
D(f_{H_n}||f_{\text{app}}) = \int_0^\infty  {{f_{H_n}}(x)  \log  {\frac{{{f_{H_n}}(x)}}{{{f_{\text{app}}}(x)}}} {\rmd}x},
\end{equation}
where \({f_{H_n}}(x)\) is defined in~\eqref{eq888} and \({f_{\text{app}}}(x)\) is the gamma approximation of \({f_{H_n}}(x)\) with the parameters \(k'\) and \(\theta '\). Fig.~\ref{fig1} illustrates the relative entropy versus \(N\) for different \(\varepsilon\). For comparison, we also plot the relative entropy of  Student's $t$-distribution with \(N\) degrees of freedom and its normal approximation (``ref'' in the legend).  As \(N\to\infty\), Student's $t$-distribution approaches the normal distribution~\cite{gb13}, but it is very accurate even for $N > 10$.  As can be seen from the figure, $D(f_{H_n}||f_{\text{app}})$ is smaller than the the relative entropy of Student's $t$-distribution and its normal approximation for reasonably large values of $N$.  This brief analysis suggests that the accuracy of the approximation resulting from the rounding operation and the use of the gamma distribution is quite high.

\subsection{One-bit Phase Adjustment}
We now treat the case where each reflector is able to adjust the phase of the incident wave by $180\degree$ only (or leave it unaltered).  Eq.~\eqref{2.17.5} can be rewritten as
\begin{equation}\small
\begin{split}
    G   &= \sum\limits_{n = 1}^N {\left| {{h_{1n}}} \right|\left| {{h_{2n}}} \right|{e^{j{\phi _n}}}}  + \left| {{h_{sd}}} \right|\\
        &= \sum\limits_{n = 1}^N {\left| {{h_{1n}}} \right|\left| {{h_{2n}}} \right|\cos ({\phi _n})}   \\
        &~~~+ j\sum\limits_{n = 1}^N {\left| {{h_{1n}}} \right|\left| {{h_{2n}}} \right|\sin ({\phi _n}) + \left| {{h_{sd}}} \right|} \\
        &= \sum\limits_{n = 1}^N {{X_n}}  + j\sum\limits_{n = 1}^N {{Y_n}}  + \left| {{h_{sd}}} \right|  \\
        &= {X} + j{Y} + R.
\end{split}\label{n1}
\end{equation}
As noted earlier, \(\phi _n\) is uniformly distributed on the interval $[-\pi/2,\pi/2]$.  As a result, the PDF of \({\cos ({\phi _n})}\) is 
\begin{equation}\small
{f_{\cos ({\phi _n})}}(x) = \frac{2}{{\pi \sqrt {1 - {x^2}} }},\quad 0\leq x \leq 1.
\label{2.8.2}
\end{equation}
We can use~\eqref{eq888} and~\eqref{2.8.2} to obtain the PDF of \({X_n}\):
\begin{equation}\label{2.7.2}\small
\begin{split}
    f_{X_n}(z) &= \int_0^1 \frac{1}{x} f_{\cos (\phi_n)}(x)f_{H_n}\lrp*{\frac{z}{x}}\rmd x \\
    &= \exp ( - z),\qquad z \geq 0.
\end{split}
\end{equation}
Similarly, the PDF of \({Y_n}\) can be calculated to be
\begin{equation}\small
    {f_{{Y_n}}}(y) = \frac{1}{2}\exp ( - \left| y \right|),\qquad y \in \mathbb R.
\label{2.7.3}
\end{equation}
Since \(X_n\) and \(Y_n\) are i.i.d.\@ for all $n$, the PDF of \(X\) and \(Y\) can be calculated to be~\cite[ch.~7]{b11}~\cite[ch.~2]{b12}
\begin{equation}\small
    {f_X}(x) = \frac{{\exp ( - x){x^{N - 1}}}}{{(N - 1)!}},\qquad x\geq 0
\label{2.7.4}
\end{equation}
and
\begin{equation}\small
\begin{split}
    {f_Y}(y) = \frac{{\exp \left( { - \left| y \right|} \right)}}{{{2^N}(N - 1)!}}  
        \sum\limits_{m = 0}^{N - 1} {\frac{{(N - 1 + m)!{{\left| y \right|}^{N - 1 - m}}}}{{{2^m}m!(N - 1 - m)!}}},\qquad y \in \mathbb R.
\end{split}\label{2.7.5}
\end{equation} 
We are now in a position to state an approximation for the CDF of \({\left| G \right|^2}\); the outage probability follows from the relation
\begin{equation}\small
    {P^{(2)}_{out}} = {F_{{{\left| G \right|}^2}}}\left( {\frac{{{\gamma _{\text{th}}}}}{{{\gamma _t}}}} \right).
\label{out222}
\end{equation}

\begin{proposition}[CDF of $|G|^2$]
    The CDF of \({\left| G \right|^2}\) satisfies the approximation
    \begin{equation}\small
    \begin{split}
       F_{|G|^2}(t) \approx& \int_0^t f_{Y^2}(y) F_X\left( \sqrt {t - y}  \right) \rmd y - \sum\limits_{i = 0}^{N - 1}\frac{B_i}{(N-1-i)!i!}
        \\& \times \int_0^t A(t-y)m(t-y)^{N - 1 - i}f_{Y^2}(y) \rmd y
    \end{split}
    \end{equation}
    where
    \begin{align}\small
        A(u) &= \exp \left( {\frac{{\sigma _d^2 - 2\sqrt {u} }}{2}} \right) \nonumber\\
        m(u) &= \sqrt {u}  - \sigma _d^2 \nonumber\\
        {B_i} &=\small{ {2^{\frac{{i - 1}}{2}}}\sigma _d^{i + 1}\left( {\Gamma \left( {\frac{{i + 1}}{2},\frac{{{m^2}}}{{2\sigma _d^2}}} \right) - \Gamma \left( {\frac{{i + 1}}{2},\frac{{\sigma _d^2}}{2}} \right)} \right) \nonumber }\\
        {F_X}( u ) &= 1 - \sum\limits_{n = 0}^{N - 1} {\frac{1}{{n!}}{u^n e^{ - u }}} \nonumber\\
        \text{and}\qquad& \nonumber \\
        {f_{{Y^2}}}(y) &= \frac{1}{{\sqrt y }}{f_Y}(\sqrt y ). \nonumber
    \end{align}
\end{proposition}
\begin{IEEEproof}
    \({F_{{{| G |}^2}}}(t)\) can be written as 
    \begin{equation}\small
        {F_{{{\left| G \right|}^2}}}(t) \approx \int_0^t {{f_{{Y^2}}}(y){\mathbb{P}}\left( {X + R < \sqrt {t - y} } \right)\rmd y} .
    \end{equation}
    Proposition 2 can be concluded by following a similar procedure as outlined in the proof of Proposition 1.
\end{IEEEproof}

Note that the approximation in Proposition 2 does not arise from the use of a moment-matched gamma distribution, but rather from the assumption that \(X\) and \(Y\) are independent.  It can be proved that \({{X_n}}\) and \({{Y_n}}\) are uncorrelated.  Indeed, when the PDF of \({{\phi _n}}\) is even, we have 
\begin{equation}\small
    {\rm E}[\cos\varphi_n \sin\varphi_n] = \frac 1 2 {\rm E}[\sin 2\varphi_n] = 0 = {\rm E}[\cos\varphi_n]{\rm E}[\sin\varphi_n]
\end{equation}
since ${\rm E}[\sin\varphi_n] = 0$.  The relative entropy of the exact distribution and the approximate distribution is just the mutual information between the random variables $X_n$ and $Y_n$:
\begin{equation}\small
    I(X_n;Y_n) = h(X_n) + h(Y_n) - h(X_n,Y_n)
\end{equation}
where $h(\cdot)$ represents the differential entropy here. The joint PDF of $(X_n,Y_n)$ can be obtained by performing a transformation of variables:
\begin{equation}\small
\begin{split}
    {f_{{X_n},{Y_n}}}(x,y) &= \det(J(x,y)) f_{H_n}(a(x,y))f_\varphi(\phi(x,y))\\
    &= \frac{1}{\pi }K_0(\sqrt {{x^2} + {y^2}} )
\end{split}
\end{equation}
where \(J(x,y) = \partial (a,\phi)/\partial (x,y)\) is the Jacobian matrix for the transformation \(\{x = a\cos \phi,\, y = a\sin \phi\}\).

By computing numerically, we have that \({I(X_n;Y_n)} \approx 0.04441\). The small value indicates the dependence between $X_n$ and $Y_n$ is small.  While this analysis applies for a single reflecting path, it is worth noting that the dependence between $X = \sum_n X_n$ and $Y = \sum_n Y_n$ will decrease as the number of elements grows.  Indeed, as the number of terms in each sum increases, central limit effects take hold and $X$ and $Y$ become approximately independent since the two random variables are (nearly) Gaussian and uncorrelated.

\section{Diversity Order}
The diversity order of the system is defined as
\begin{equation}\small
    {d}_{i} = \mathop {\lim }\limits_{{\gamma _t} \to \infty }  - \frac{{\log  {{P^{(i)}_{\text{out}}}} }}{{\log {\gamma _t}}}.
\label{d2}
\end{equation}
From the analysis presented in the previous section, it is clear that the diversity order depends on the small-argument behaviour of $F_H$ and $F_{|G|^2}$ for the systems studied herein.  Below, we conduct an asymptotic analysis of the CDFs, which leads to diversity order expressions for the cases of perfect phase alignment and one-bit phase adjustments.  

\subsection{Perfect Phase Alignment}
In the previous section, we invoked a gamma approximation to analyze the outage probability of the IRS-aided system under the assumption of perfect phase alignment.  Here, we refrain from using this approximation.  The asymptotic analysis that follows is exact.

We analyze \({F_H}(t)\) when \(t\) approaches zero. The PDF of \(H_n\) can be expanded about zero to yield \({f_{H_n}}(x) = - x\ln x + O(x)\). Performing an $N$-fold convolution and taking the leading term every time, we obtain the leading order of the PDF of \(S\):
\begin{equation}\small
{f_S}(x) = \frac{{{x^{2N - 1}}}}{{(2N - 1)!}}{\left( {\ln \frac{1}{x}} \right)^N} + o\left( {{x^{2N - 1}}{{\left( {\ln x} \right)}^N}} \right).
\label{ln_order}
\end{equation}
Eq.~\eqref{ln_order} can be proved by induction. Similarly, the PDF of \(R = \left| {{h_{sd}}} \right|\) can be written to leading order as \(f_R(x) \approx x/{{\sigma _d^2}}\). Hence, the CDF of \(H\), to leading order, is the integral of the convolution of $f_R$ and $f_S$:
\begin{equation}\small
\begin{split}
{F_H}(t) &= \int_0^t {\int_0^s {{f_S}(x){f_R}(s - x)\,{\rm{d}}x} {\rm{d}}s} \\
&= \frac{{{t^{2(N + 1)}}{{\left( {\ln \frac{1}{t}} \right)}^N}}}{{4\sigma _d^2N(N + 1)(2N + 1)(2N - 1)!}} + \epsilon(t)
\end{split}
\end{equation}
where $\epsilon(t) = o(t^{2(N+1)} (\ln t)^N)$. Based on the definition of the outage probability in~\eqref{out1} and the definition of diversity order in~\eqref{d2}, we find that the diversity order under perfect phase alignment is
\begin{equation}\small
    d_1 = N+1.
\end{equation}
This is entirely expected, since the assumption of the model we consider is that all $N+1$ spatial channels are independent.

\subsection{One-bit Phase Adjustment}
In a similar manner as was done for the case of perfect phase alignment, we start by analyzing \({F_{{{\left| G \right|}^2}}}(t)\) when \(t\) approaches zero. The PDF of \(X+R\) can be obtained by computing the convolution of the leading orders of the PDFs of \(X\) and \(R\).  Note that in this case the PDF of $X$ has the simple (exact) form given by~\eqref{2.7.4}.  Computing the convolution, retaining the leading order, and performing a transformation of variables, we arrive at the PDF of \((X+R)^2\), which is
\begin{equation}\small
{f_{{{\left( {X + R} \right)}^2}}}(x) = \frac{{{x^{\frac{N}{2}}}}}{{2\sigma _d^2(N + 1)!}} + o(x^{N/2}).
\end{equation}
Similarly, we can write the leading order term of the PDF of \(Y^2\) in the succinct form
\begin{equation}\small
{f_{{Y^2}}}(y) = \frac{{\Gamma \left( {N - \frac{1}{2}} \right)}}{{2 \Gamma(N) \sqrt{\pi y} }} + O(1).
\end{equation}
Invoking the assumption that the real and imaginary parts of the channel gain are independent, we arrive at the following approximation for the leading order of the CDF of \({{{\left| G \right|}^2}}\):
\begin{equation}\small
\begin{split}
{F_{{{\left| G \right|}^2}}}(t) &\approx \int_0^t {\int_0^{t - y} {{f_{{Y^2}}}(y){f_{{{(X + R)}^2}}}(x) \rmd x} \rmd y} \\
 &\approx \frac{{{t^{\frac{{3 + n}}{2}}}\Gamma \left( {\frac{N}{2} + 2} \right)\Gamma \left( {N - \frac{1}{2}} \right)}}{{2\sigma _d^2\Gamma \left( N \right)\Gamma \left( {N + 3} \right)\Gamma \left( {\frac{{N + 5}}{2}} \right)}}.
\end{split}
\end{equation}
The inner integral above can be evaluated directly, and the outer integral results from~\cite[eq.~3.191.1]{integral}.

From~\eqref{out222} and~\eqref{d2}, we have that
\begin{equation}\small
    d_2 = \frac{{N + 3}}{2}.
\label{order_error}
\end{equation}
This is a curious result, which requires further investigation.  It is unclear exactly how the phase quantization leads to this asymptotic behavior.  One might suspect that the independence assumption yields an erroneous result.  We argue in the next section, by using numerical simulations, that this is not the case.  It would be interesting to develop an exact asymptotic expression for the distribution of $|G|^2$ where independence between the real and imaginary parts is not assumed.  We refrain from doing so here, however, favoring a thorough numerical investigation instead.

\section{Numerical Results}
Denoting the distances of the S-to-\({\rm R}_n\), \({\rm R}_n\)-to-D, and S-to-D links as \({d_{SR}}\), \({d_{RD}}\) and \({d_{SD}}\), respectively, the path losses of the channels are calculated by using the 3GPP Urban Micro NLOS model with a carrier frequency of 5 GHz~\cite{3gpp}, which states that $\xi =  - 40.9 - 36.7{\log _{10}}(d)~\text{dB}$, where \(d\) is the distance. In simulations, we set \(\gamma_{\text{th}}=0\,\rm{dB}\).

\begin{figure}[t!]
\centerline{\includegraphics[scale=0.45]{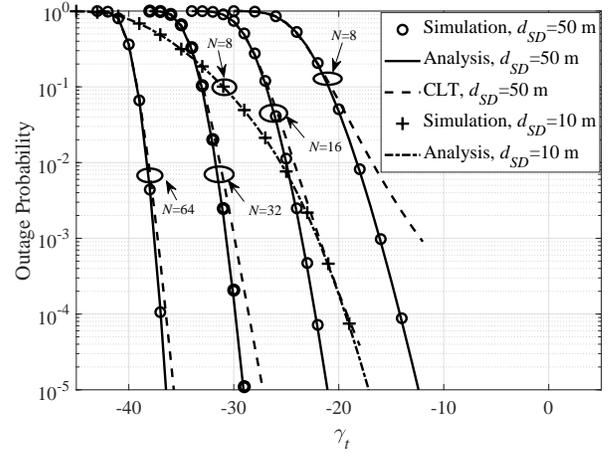}}
\caption{Outage probability versus transmit SNR without phase errors for different $N$. The distances are set to \({d_{SR}} = 40\,\rm{m}\), \({d_{RD}} = 30\,\rm{m}\).}
\label{fig2}
\end{figure}

The outage probability is plotted against the transmit SNR \({\gamma _t}\) when there are no phase errors in Fig.~\ref{fig2}. In particular, the outage probability is compared with the CLT method and Monte-Carlo (MC) simulations. We observe that the outage probability of the proposed approximation is more accurate than that based on the CLT method. Fig.~\ref{fig2} also illustrates that with an increasing number of reflecting elements, the outage probability decreases significantly for the same transmit SNR. For example, when \({d_{SD}}=50\,\rm{m}\) and \(\gamma_{t}=-25\,\rm{dB}\), the outage probability is $0.69$ for \(N=8\) and $10^{-2}$ for \(N=16\). To ascertain the influence of the direct link, we also present results for when the IRS is placed such that \({d_{SD}}=10\,\rm{m}\). Since \({d_{SD}} \ll {d_{SR}} + {d_{RD}}\), this placement leads to a strong direct link. In this case, the diversity order emerges slowly as the transmit SNR increases, as one would expect.  At high SNR, the curve becomes parallel to that corresponding to \({d_{SD}}=50\,\rm{m}\); however, a coding gain is observed in the SNR shift to the left, which is a result of the stronger direct link.

\begin{figure}[t!]
\centerline{\includegraphics[scale=0.45]{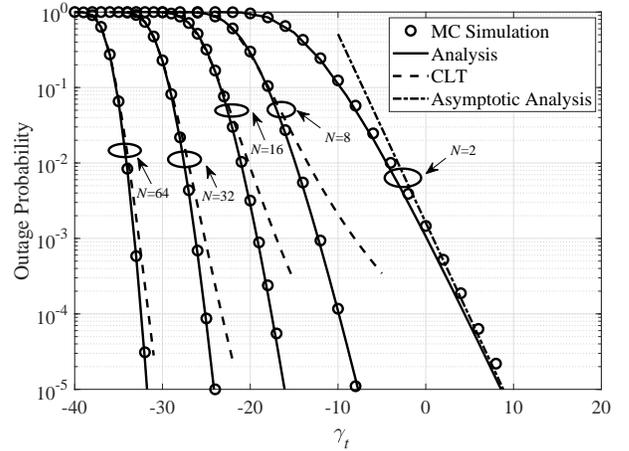}}
\caption{Outage probability versus transmit SNR with phase errors for different $N$. The distances are set to \({d_{SR}} =40\,\rm{m}\), \({d_{RD}} = 30\,\rm{m}\), \({d_{SD}} = 50\,\rm{m}\).}
\label{fig3}
\end{figure}

Fig.~\ref{fig3} illustrates the outage probability under the condition of one-bit phase adjustment. Similar to the case without phase errors, our analysis is in good agreement with the simulations. Comparing the results in Figs.~\ref{fig2} and~\ref{fig3}, it can be seen that phase errors lead to a performance loss with the gap being about $5\,\rm{dB}$.  For example, when \(N=16\), the transmit SNRs required to achieve an outage probability of $10^{-2}$ are $-25\,\rm{dB}$ and $-20\,\rm{dB}$ for perfect and imperfect phase alignment, respectively. The asymptotic outage probability is plotted for $N=2$ in this example as well. This curve is included to illustrate that the approximate asymptotic analysis resulting from the independence assumption is reasonably accurate.  Hence, we can be fairly confident that the diversity order is indeed $(N+3)/2$ in this case.

\section{Conclusions}
In this paper, the performance of IRS-assisted communication systems was analyzed.  It was shown that the existence of a direct link slows the emergence of the diversity slope, and that one can expect a penalty of approximately 5 dB when a one-bit phase adjustment is made at each reflective element.  We also demonstrated the importance of \textit{not} using a CLT approximation to study high-SNR behavior.  Through a more accurate asymptotic analysis (though still inexact), we conjectured that the diversity order of a one-bit phase adjustment system is $(N+3)/2$.  This result requires further investigation, and an exact understanding of the performance of systems that employ a $b$-bit phase adjustment at each element constitutes an important question for further study.  We plan to address this problem in future work.


\appendix
We begin with the following two lemmas.


\begin{lemma}
    Let $Q$ be a non-negative random variable, and let $R$ be Rayleigh distributed with parameter $\sigma$ such that $Q$ and $R$ are independent.  Let $Z = Q + R$.  Then the distribution function of $Z$ satisfies the equation
    \begin{equation}\small
        F_Z(z) = F_Q(z) - \int_0^z f_Q(q)\exp \left(  - \frac{(z - q)^2}{2\sigma^2} \right)\rmd q.
    \end{equation}
\end{lemma}
\begin{IEEEproof}
    The proof follows by noting that
    \begin{equation}\small
       F_Z(z) = \int_0^z f_Q(q){\mathbb{P}}\left( R^2 < (z - q)^2 \right) \rmd q
    \end{equation}
    and that $R^2$ is exponentially distributed with mean $2\sigma^2$.
\end{IEEEproof}

\begin{lemma}
    Let $I$ be a positive integer, and let $a$ and $b$ be two real numbers. Then the following relation is true:
    \begin{equation}\small
    \begin{split}
        \int_0^t {{x^I}\exp ( - ax)\exp \left( { - b{{(t - x)}^2}} \right){\rm{d}}x}  = \frac{1}{2}\exp \left( {\frac{{{a^2} - 4abt}}{{4b}}} \right)\\
        \times\sum\limits_{i = 0}^I {\binom{I}{i}{m^{I - i}}\left( {{b^{\frac{{ - 1 - i}}{2}}}{E_i}} \right)}
    \end{split}
    \end{equation}
    where $m = (2bt - a)/(2b)$ and
    \begin{equation}\small
        {E_i} = \Gamma \left( {\frac{{i + 1}}{2},b{m}} \right) - \Gamma \left( {\frac{{i + 1}}{2},b{{(t - m)}^2}} \right).
    \end{equation}
\end{lemma}
\begin{IEEEproof}
    The proof follows by noting that 
    \begin{equation}\small
    \begin{split}
        \int_0^t {{x^I}\exp ( - ax)\exp \left( { - b{{(t - x)}^2}} \right){\rm{d}}x}  =\exp \left( {\frac{{{a^2} - 4abt}}{{4b}}} \right)\\
        \times\int_{ - m}^{t - m} {{{(n + m)}^I}\exp \left( { - b{n^2}} \right){\rm{d}}n}, 
    \end{split}
    \end{equation}
    where $n = x - m$, and the integral
    \begin{equation}\small
        \int {{n^i}\exp \left( { - b{n^2}} \right){\rm d}n}  =  - \frac{1}{2}{b^{\frac{{ - 1 - i}}{2}}}\Gamma \left( {\frac{{i + 1}}{2},b{n^2}} \right).
    \end{equation}
\end{IEEEproof}
Since $H = S + R$ in Proposition 1, the CDF of $H$ satisfies (by Lemma 1)
\begin{equation}\small
    {F_H}(t) = {F_S}(t) - \int_0^t {{f_S}} (r)\exp \left( { - \frac{{{{(t - r)}^2}}}{{2{\sigma_d ^2}}}} \right)\rmd r.
    \label{appenP}
\end{equation}
The integral in \eqref{appenP} can be calculated with Lemma 2.


\bibliographystyle{IEEEtran}
\bibliography{irsgc}

\begin{thebibliography}{10}
\providecommand{\url}[1]{#1}
\csname url@samestyle\endcsname
\providecommand{\newblock}{\relax}
\providecommand{\bibinfo}[2]{#2}
\providecommand{\BIBentrySTDinterwordspacing}{\spaceskip=0pt\relax}
\providecommand{\BIBentryALTinterwordstretchfactor}{4}
\providecommand{\BIBentryALTinterwordspacing}{\spaceskip=\fontdimen2\font plus
\BIBentryALTinterwordstretchfactor\fontdimen3\font minus
  \fontdimen4\font\relax}
\providecommand{\BIBforeignlanguage}[2]{{%
\expandafter\ifx\csname l@#1\endcsname\relax
\typeout{** WARNING: IEEEtran.bst: No hyphenation pattern has been}%
\typeout{** loaded for the language `#1'. Using the pattern for}%
\typeout{** the default language instead.}%
\else
\language=\csname l@#1\endcsname
\fi
#2}}
\providecommand{\BIBdecl}{\relax}
\BIBdecl

\bibitem{b4}
C.~Liaskos, S.~Nie \emph{et~al.}, ``A new wireless communication paradigm
  through software-controlled metasurfaces,'' \emph{IEEE Commun. Mag.},
  vol.~56, no.~9, pp. 162--169, 2018.

\bibitem{b1}
C.~{Liaskos}, S.~{Nie} \emph{et~al.}, ``Realizing wireless communication
  through software-defined hypersurface environments,'' in \emph{Proc. IEEE
  WoWMoM}, Chania, Greece, Jun. 2018, pp. 14--15.

\bibitem{b3}
L.~{Subrt} and P.~{Pechac}, ``Controlling propagation environments using
  intelligent walls,'' in \emph{Proc. EUCAP}, Prague, Czech Republic, Mar.
  2012, pp. 1--5.

\bibitem{zr0}
Q.~Wu and R.~Zhang, ``Intelligent reflecting surface enhanced wireless network:
  Joint active and passive beamforming design,'' in \emph{Proc. IEEE GLOBECOM},
  Abu Dhabi, United Arab Emirates, Dec. 2018, pp. 1--6.

\bibitem{RS19}
X.~Yu, D.~Xu \emph{et~al.}, ``Robust and secure wireless communications via
  intelligent reflecting surfaces,'' \emph{arXiv preprint arXiv:1912.01497},
  2019.

\bibitem{b2}
F.~{Liu}, A.~{Pitilakis} \emph{et~al.}, ``Programmable metasurfaces: State of
  the art and prospects,'' in \emph{Proc. IEEE ISCAS}, Florence, Italy, May.
  2018, pp. 1--5.

\bibitem{bm}
E.~{Basar}, ``Transmission through large intelligent surfaces: A new frontier
  in wireless communications,'' in \emph{Proc. EuCNC}, Valencia, Spain, Jun.
  2019, pp. 112--117.

\bibitem{zr}
Q.~Wu and R.~Zhang, ``Beamforming optimization for intelligent reflecting
  surface with discrete phase shifts,'' in \emph{Proc. IEEE ICASSP}, Brighton,
  United Kingdom, May. 2019, pp. 7830--7833.

\bibitem{b5}
C.~Huang, A.~Zappone \emph{et~al.}, ``Reconfigurable intelligent surfaces for
  energy efficiency in wireless communication,'' \emph{IEEE Trans. Wireless
  Commun.}, vol.~18, no.~8, pp. 4157--4170, 2019.

\bibitem{b6}
E.~Bj{\"o}rnson, {\"O}.~{\"O}zdogan, and E.~G. Larsson, ``Intelligent
  reflecting surface vs. decode-and-forward: How large surfaces are needed to
  beat relaying?'' \emph{IEEE Wireless Commun. Lett.}, 2019.

\bibitem{b7}
D.~Kudathanthirige, D.~Gunasinghe, and G.~Amarasuriya, ``Performance analysis
  of intelligent reflective surfaces for wireless communication,'' \emph{arXiv
  preprint arXiv:2002.05603}, 2020.

\bibitem{b8}
M.-A. Badiu and J.~P. Coon, ``Communication through a large reflecting surface
  with phase errors,'' \emph{IEEE Wireless Commun. Lett.}, 2019.

\bibitem{b9}
S.~Atapattu, R.~Fan \emph{et~al.}, ``Reconfigurable intelligent surface
  assisted two-way communications: Performance analysis and optimization,''
  \emph{arXiv preprint arXiv:2001.07907}, 2020.

\bibitem{b10}
J.~Salo, H.~M. El-Sallabi, and P.~Vainikainen, ``The distribution of the
  product of independent rayleigh random variables,'' \emph{IEEE Trans.
  Antennas Propag.}, vol.~54, no.~2, pp. 639--643, 2006.

\bibitem{gb13}
B.~Grigelionis, \emph{Student's t-distribution and related stochastic
  processes}.\hskip 1em plus 0.5em minus 0.4em\relax Springer, 2013.

\bibitem{b11}
J.~L. Snell, \emph{Introduction to probability}.\hskip 1em plus 0.5em minus
  0.4em\relax Random House New York, 1988.

\bibitem{b12}
S.~Kotz, T.~Kozubowski, and K.~Podgorski, \emph{The Laplace distribution and
  generalizations: a revisit with applications to communications, economics,
  engineering, and finance}.\hskip 1em plus 0.5em minus 0.4em\relax Springer
  Science \& Business Media, 2012.

\bibitem{integral}
I.~S. Gradshteyn and I.~M. Ryzhik, \emph{Table of integrals, series, and
  products}.\hskip 1em plus 0.5em minus 0.4em\relax Academic press, 2014.

\bibitem{3gpp}
\emph{Further advancements for E-UTRA physical layer aspects (Release
  9)}.\hskip 1em plus 0.5em minus 0.4em\relax 3GPP TS 36.814, Mar. 2017.

\end{thebibliography}

\end{document}